\documentclass[aps,prd,preprintnumbers,showpacs,nofootinbib,onecolumn]{revtex4-1}

\usepackage{amsmath}

\begin{document}
\title{New torsion black hole solutions in Poincar\'e gauge theory}

\author{Jose A. R. Cembranos$^{1,2,}$\footnote{E-mail: cembra@fis.ucm.es},
and Jorge Gigante Valcarcel$^{1,}$\footnote{E-mail: jorgegigante@ucm.es}
}

\affiliation{$^{1}$Departamento de  F\'{\i}sica Te\'orica I, Universidad Complutense de Madrid, E-28040 Madrid, Spain\\
$^{2}$Departamento de  F\'{\i}sica, Universidade de Lisboa,
P-1749-016 Lisbon, Portugal
}

\pacs{04.70.Bw, 04.40.-b, 04.20.Jb, 11.10.Lm, 04.50.Kd, 04.50.-h}


%
%
%

\begin{abstract}
We derive a new exact static and spherically symmetric vacuum solution in the framework of the Poincar\'e gauge field theory with dynamical massless torsion. This theory is built in such a form that allows to recover General Relativity when the first Bianchi identity of the model is fulfilled by the total curvature. The solution shows a Reissner-Nordstr\"{o}m type geometry with a Coulomb-like curvature provided by the torsion field. It is also shown the existence of a generalized Reissner-Nordstr\"{o}m-de Sitter solution when additional electromagnetic fields and/or a cosmological constant are coupled to gravity.
\end{abstract}

\maketitle

\section{Introduction}

General Relativity (GR) is the most successful and accurate theory of classical gravity from the last century. Its outstanding description of the gravitational interaction as a purely geometrical effect of the space-time together with a large number of experimental evidences has exalted it as the fundamental theoretical basis for modern astrophysics and cosmology \cite{Will}. Even nowadays, its elemental foundations and further implications are continually being reviewed and tested, as in the case of the recent discovery of gravitational waves from a binary black hole system \cite{Abbott}. Nevertheless, extensions of GR have always attracted much attention due to the deep related fundamental questions and open questions still unsolved by the theory, as the formulation of a consistent quantum field approach to gravity, the understanding of space-time singularities or the nature of dark energy, dark matter or inflation in the very early Universe \cite{Cap-Lau,DeFelice,Berti,Avelino}.

Another open issue consists in providing correctly the foundations of the angular momentum of gravitating sources and its suitable conservation laws in presence of a dynamical space-time within the same framework. Specifically, the intrinsic angular momentum of matter must be represented by a spin density tensor and therefore it may be expected to have it associated with a fundamental geometrical quantity. However, in standard GR, it does not couple to any distinctive geometrical property, so it is analysed possible modifications of the theory according to these lines.

In this sense, Poincar\'e Gauge (PG) theory provides the most elegant and promising extension of GR, in the framework of a Riemann-Cartan (RC) manifold (i.e. a manifold endowed with curvature and torsion), in order to couple the spin of particles to the torsion of the space-time \cite{Kibble,Hehl}. Indeed, within this model, both energy-momentum and spin tensors of gravitating matter act as sources of the interaction. In addition, the role of torsion depends on the order of the field strength tensors included in the Lagrangian: whereas the full linear case involves a non-propagating torsion (i.e. tied to spinning material sources), higher order corrections describe a Lagrangian with dynamical torsion 
\cite{Obukhov,Obukhovv}.

Furthermore, the vacuum structure of the space-time also differs depending on this critical role, especially when a certain class of PG models provides the existence of propagating torsion modes in vacuum. Specifically, Birkhoff's theorem establishing that the only vacuum solution with spherical symmetry is the Schwarzschild solution, is satisfied only in certain cases of the PG theory \cite{Nev,Rauch-Nieh}. In this work, we consider a particular PG theory described by a Lagrangian of first and second order in the curvature terms, which reduces to ordinary GR when torsion satisfies a general condition connected to the first Bianchi identity. Only in such a case, it loses its physical relevance. It is shown that within this framework, the Birkhoff's theorem is not satisfied and a new analytical SO(3) spherically symmetric and static vacuum solution with dynamical torsion emerges. This solution describes a Reissner-Nordstr\"{o}m type configuration characterized exclusively for its mass and the torsion field contribution, in analogy to the electric charge in Maxwell's theory. Thus, by this contribution of the torsion field to the space-time geometry, neither other physical sources nor electromagnetic fields are necessary to generate this type of solutions. On the other hand, we also stress that it is always possible to find a generalized Reissner-Nordstr\"{o}m-de Sitter solution endowed with both electric and magnetic charges, as well as with a cosmological constant within this construction. Finally, the equations of motion for a general test particle in such a space-time are obtained from the respective conservation law of the energy-momentum tensor of matter.

This work is organized as follows. First, in Section II, we briefly present the general mathematical foundations of PG theory paying spetial attention to our model. Field equations and analyses of general solutions beyond the Birkhoff's theorem for GR and different classes of PG theories are shown in Section III. Our new analytical solution within this framework, as well as its natural generalization to include external Coulomb electric and magnetic fields with a non-vanishing cosmological constant are presented and analysed in Section IV. In Section V, we obtain from the general conservation law of the energy-momentum tensor, the equations of motion for a test particle belonging to a RC manifold connected to our model. Finally, we present the conclusions of our work in Section VI. A general demonstration for the conservation law of the energy-momentum tensor is also presented in Appendix A.

Before proceeding to the main discussion and general results, we briefly introduce the notation and physical units to be used throughout this article. Latin $a,b$ and greek $\mu, \nu$ indices refer to anholonomic and coordinate basis, respectively. We use notation with tilde for magnitudes including torsion (i.e. defined within a RC manifold) and without tilde for torsionless objects. Finally, we will use Planck units ($G=c=\hbar=1)$.

\section{Quadratic Poincar\'e gauge gravity model}

A model of PG gravity requires gauging the external degrees of freedom consisting of rotations and translations, which are represented by the Poincar\'e group $ISO (1,3)$. Therefore, a gauge connection containing two principal independent variables is introduced in order to describe the gravitational field. These quantities constitute the gauge potentials related to the generators of translations and local Lorentz rotations, respectively:

\begin{equation}A_{\mu}=e^{a}\,_{\mu}P_{a}+\omega^{a b}\,_{\mu}J_{a b}\,,\end{equation}
where $e^{a}\,_{\mu}$ is the vierbein field and $\omega^{a b}\,_{\mu}$ the spin connection, which satisfy the following relations with the metric $g$ and the affine connection $\tilde{\Gamma}$ within the RC manifold \cite{Yepez}:

\begin{equation}
g_{\mu \nu}=e^{a}\,_{\mu}\,e^{b}\,_{\nu}\,\eta_{a b}\,,
\end{equation}

\begin{equation}
\omega^{a b}\,_{\mu}=e^{a}\,_{\lambda}\,e^{b \rho}\,\tilde{\Gamma}^{\lambda}\,_{\rho \mu}+e^{a}\,_{\lambda}\,\partial_{\mu}\,e^{b \lambda}\,.
\end{equation}

Note that in a RC manifold the affine connection constitutes a metric-compatible connection (i.e. $\tilde{\nabla}_{\lambda}\,g_{\mu\nu} = 0$). Moreover, it can split into the Levi-Civita connection and the so called contortion tensor in the following way:

\begin{equation}
\tilde{\Gamma}^{\lambda}\,_{\mu \nu} = {\Gamma}^{\lambda}\,_{\mu \nu} + K^{\lambda}\,_{\mu \nu}\,.
\end{equation}

Additionally, $P_{a}$ are the generators of the space-time translations and $J_{a b}$ the generators of the space-time rotations, which satisfy the following commutative relations:

\begin{equation}
\left[P_{a},P_{b}\right]=0\,,
\end{equation}

\begin{equation}
\left[P_{a},J_{bc}\right]=i\,\eta_{a[b}\,P_{c]}\,,
\end{equation}

\begin{equation}
\left[J_{ab},J_{cd}\right]=\frac{i}{2}\,\left(\eta_{ad}\,J_{bc}+\eta_{cb}\,J_{ad}-\eta_{db}\,J_{ac}-\eta_{ac}\,J_{bd}\right)\,.
\end{equation}

Then, the corresponding $ISO (1,3)$ gauge field strength tensor defined by $F_{\mu \nu}=\partial_{\mu}A_{\nu}-\partial_{\nu}A_{\mu}-i[A_{\mu},A_{\nu}]$ takes the form:

\begin{equation}F_{\mu \nu}=F^{a}\,_{\mu \nu}P_{a}+F^{a b}\,_{\mu \nu}J_{a b}\,,\end{equation}
with $F^{a}\,_{\mu \nu}= \partial_{\mu}e^{a}\,_{\nu}-\partial_{\nu}e^{a}
\,_{\mu}+\omega^{a b}\,_{\mu}e\,_{b \nu}-\omega^{a b}\,_{\nu}\,e_{b \mu}\,$,
and $F^{a b}\,_{\mu \nu}=\partial_{\mu}\omega^{ab}\,_{\nu}
-\partial_{\nu}\omega^{a b}\,_{\mu}+\omega^{a c}\,_{\nu}
\,\omega^{b}\,_{c\mu}-\omega^{a c}\,_{\mu}\,\omega^{b}\,_{c \nu}\,$.
\\

As in the case of other known gauge theories, the field strength tensor characterizes the properties of the gravitational interaction, that in the PG framework are potentially modified by the presence of torsion. In particular, it is related to the torsion and the curvature of the space-time as follows:

\begin{equation}F^{a}\,_{\mu \nu}=e^{a}\,_{\lambda}\,T^{\lambda}\,_{\nu \mu}\,,\end{equation}

\begin{equation}F^{a b}\,_{\mu \nu}=e^{a}\,_{\lambda}e^{b}\,_{\rho}\,\tilde{R}^{\lambda \rho}\,_{\mu \nu}\,,\end{equation}
where $T^{\lambda}\,_{\mu \nu}$ and $\tilde{R}^{\lambda \rho}\,_{\mu \nu}$ are the components of the torsion and the curvature tensor respectively:

\begin{equation}T^{\lambda}\,_{\mu \nu}=2\tilde{\Gamma}^{\lambda}\,_{[\mu \nu]}\,,\end{equation}

\begin{equation}\tilde{R}^{\lambda}\,_{\rho \mu \nu}=\partial_{\mu}\tilde{\Gamma}^{\lambda}\,_{\rho \nu}-\partial_{\nu}\tilde{\Gamma}^{\lambda}\,_{\rho \mu}+\tilde{\Gamma}^{\lambda}\,_{\sigma \mu}\tilde{\Gamma}^{\sigma}\,_{\rho \nu}-\tilde{\Gamma}^{\lambda}\,_{\sigma \nu}\tilde{\Gamma}^{\sigma}\,_{\rho \mu}\,.\end{equation}

These components modify the commutative relations of the covariant derivatives for a general vector field $v^{\lambda}$ over a RC manifold in the following way:

\begin{equation}[\tilde{\nabla}_{\mu},\tilde{\nabla}_{\nu}]\,v^{\lambda}=\tilde{R}^{\lambda}\,_{\rho \mu \nu}\,v^{\rho}+T^{\rho}\,_{\mu \nu}\,\tilde{\nabla}_{\rho}v^{\lambda}\,,
\end{equation}
with $\tilde{\nabla}_{\mu}\,v^{\lambda}=\partial_{\mu}\,v^{\lambda}+\tilde{\Gamma}^{\lambda}\,_{\rho \mu}\,v^{\rho}$.

Hence, whereas curvature is related to the rotation of a vector
along an infinitesimal path over the space-time, torsion is related to the translation and it has deep geometrical implications, such as breaking infinitesimal parallelograms on the manifold \cite{Sab-Gas}. Furthermore, the RC manifold may be regarded as an effective geometrical construction arising from a microscopic structure endowed with dislocation defects, which are described by torsion in the limit where they form a continuous distribution \cite{Sab-Siv,Rubiera}. In this sense, it is expected that the field strength tensor defined within this RC manifold gives rise to the pattern of dislocations density in terms of a dynamical torsion (i.e. even in the absence of matter fields).

In addition, both curvature and torsion tensors can also be classified by the decomposition into their irreducible parts under the Lorentz group \cite{Gambini,Hay-Shi}. Especially, torsion can be divided into three irreducible components given by distinct contributions: a trace vector, an axial vector and a traceless and also pseudotraceless tensor. From a phenomenological point of view, this sort of geometrical classification can be associated with a large number of physically relevant situations, such as the coupling between the Dirac fields and the totally antisymmetric part of the torsion or the vanishing of its tensorial modes in a spatially homogeneous and isotropic universe, as it is assumed by the cosmological principle (see \cite{Capozziello} for a more detailed account and alternative classifications). However, there exist more complex systems that require the non-vanishing of the rest of the modes, such as the given by a general static and spherically symmetric space-time, which is deeply considered in this work.

In the basic version of the PG theory, the presence of torsion is sourced by the spin of matter, so that it introduces new independent characteristics from the standard theory and it achieves a dynamical role defining an invariant Lagrangian quadratic in the field strength tensors. In this work, we focus on a PG model whose second order contributions are only due to the existence of this kind of non-vanishing and also massless torsion:

\begin{eqnarray}
S &=& \; \frac{1}{16 \pi}\int d^4x \sqrt{-g}
\Bigl[
\mathcal{L}_{m}-R-\frac{1}{4}\left(d_{1}+d_{2}\right)\tilde{R}^2-\frac{1}{4}\left(d_{1}+d_{2}+4c_{1}+2c_{2}\right)\tilde{R}_{\lambda \rho \mu \nu}\tilde{R}^{\mu \nu \lambda \rho}
\Bigr.
\nonumber\\
& &
\Bigl.\;\;\;\;\;\;\;\;\;\;\;\;\;\;\;\;\;\;\;\;\;\;\;\;\;\;\;\;\;\;\;
+c_{1}\tilde{R}_{\lambda \rho \mu \nu}\tilde{R}^{\lambda \rho \mu \nu}+c_{2}\tilde{R}_{\lambda \rho \mu \nu}\tilde{R}^{\lambda \mu \rho \nu}
+d_{1}\tilde{R}_{\mu \nu}\tilde{R}^{\mu\nu}+d_{2}\tilde{R}_{\mu\nu}\tilde{R}^{\nu\mu}
\Bigr]\,,
\label{actioneq}
\end{eqnarray}
where $c_{1},c_{2},d_{1}$ and $d_{2}$ are four constant parameters. Note that in order to construct the Expression \eqref{actioneq}, we can use the identity $\tilde{R}=R-2\nabla_{\lambda}T^{\rho \lambda}\,_{\rho}+\frac{1}{4}T_{\lambda \mu \nu}T^{\lambda \mu \nu}+\frac{1}{2}T_{\lambda \mu \nu}T^{\mu \lambda \nu}-T^{\mu}\,_{\mu \lambda}T^{\nu}\,_{\nu}\,^{\lambda}$, which allows to rewrite the general PG Lagrangian with massless torsion in terms of the torsionless Einstein-Hilbert Lagrangian.

In the elementary case where torsion does not propagate, all these constants vanish and the action leads to the standard Einstein theory. However, as it is remarked above, we are interested in the presence of higher order curvature terms in the action because in such a case, torsion becomes dynamical. Furthermore, in the limit where the first Bianchi identity of GR still holds for the total curvature (i.e. $\tilde{R}^{\lambda}\,_{[\mu \nu \rho]}=0$
\footnote{The symmetric and antisymmetric parts of a generic covariant tensor $A_{a_{1} ... a_{q}}$ are denoted by parenthesis and brackets, respectively:
\begin{equation}
A_{a_{1} ... a_{q}} = A_{(a_{1} ... a_{q})} + A_{[a_{1} ... a_{q}]}\,,
\end{equation}
with:
\begin{equation}
A_{(a_{1} ... a_{q})} = \frac{1}{q!}\sum_{\pi} A_{a_{\pi (1)} ... a_{\pi (q)}}\,,
\end{equation}
and
\begin{equation}
A_{[a_{1} ... a_{q}]} = \frac{1}{q!}\sum_{\pi} \delta_{\pi} A_{a_{\pi (1)} ... a_{\pi (q)}}\,,
\end{equation}
where the sum is taken over all permutations $\pi$ of $1, ... , q$ and $\delta_{\pi}$ is $+1$ for even permutations and $-1$ for odd permutations.}
), then the Lagrangian leads to the sum of the Einstein-Hilbert Lagrangian and the Gauss-Bonnet term. As it is well known, the latter is a topological invariant in the four dimensional case, so it does not contribute to the field equations and the theory coincides locally with GR.

According to the first Bianchi identity in a RC space-time \cite{Ortin}:

\begin{equation}\tilde{R}^{\lambda}\,_{[\mu \nu \rho]}+\tilde{\nabla}_{[\mu}T^{\lambda}\,_{\nu \rho]}+T^{\sigma}\,_{[\mu \nu}\,T^{\lambda}\,_{\rho] \sigma}=0\,,\end{equation}
the Expression (\ref{actioneq}) reduces to the regular gravity action when $\tilde{\nabla}_{[\mu}T^{\lambda}\,_{\nu \rho]}+T^{\sigma}\,_{[\mu \nu}\,T^{\lambda}\,_{\rho] \sigma}=0$. Note that this expression does not imply the vanishing of the torsion tensor, but a less constraining condition fulfilled by this quantity for recovering GR.

\section{Field equations}

In order to derive the field equations, we may simplify the expression above without loss of generality applying the Gauss-Bonnet theorem in RC spaces \cite{Nieh,Hayashi}. Indeed, the following term is a total derivative of a certain
vector $V^\mu$:

\begin{equation}\sqrt{- g}\,\left(\tilde{R}^{2}+\tilde{R}_{\lambda \rho \mu \nu}\tilde{R}^{\mu \nu \lambda \rho}-4\tilde{R}_{\mu \nu}\tilde{R}^{\nu \mu}\right)=\partial_\mu V^\mu\,.\end{equation}
Then (\ref{actioneq}) is locally equivalent to the following action:

\begin{equation}
\label{actioneqq}
S=\frac{1}{16 \pi}\int d^4x\sqrt{-g}
\left[
\mathcal{L}_{m}-R-\frac{1}{2}\left(2c_{1}+c_{2}\right)\tilde{R}_{\lambda \rho \mu \nu}\tilde{R}^{\mu \nu \lambda \rho}+c_{1}\tilde{R}_{\lambda \rho \mu \nu}\tilde{R}^{\lambda \rho \mu \nu}+c_{2}\tilde{R}_{\lambda \rho \mu \nu}\tilde{R}^{\lambda \mu \rho \nu}+d_{1}\tilde{R}_{\mu \nu}\left(\tilde{R}^{\mu \nu}-\tilde{R}^{\nu \mu}\right)
\right]\,.
\end{equation}

In the absence of matter, i.e. $\mathcal{L}_{m}=0$, Birkhoff's theorem is satisfied only in certain cases of the PG theory \cite{Nev,Rauch-Nieh}. We observe that our particular PG model does not generally satisfy this theorem, so the analysis of new static and spherically symmetric vacuum solutions to the field equations is necessary.

Before computing the vacuum equations, we define the following geometric quantities:

\begin{equation}G_{\mu}\,^{\nu} = R_{\mu}\,^{\nu}-\frac{R}{2}\delta_{\mu}\,^{\nu}\,,\end{equation}

\begin{equation}T1_{\mu}\,^{\nu} = \tilde{R}_{\lambda \rho \mu \sigma}\tilde{R}^{\lambda \rho \nu \sigma}-\frac{1}{4}\delta_{\mu}\,^{\nu}\tilde{R}_{\lambda \rho \tau \sigma}\tilde{R}^{\lambda \rho \tau \sigma}\,,\end{equation}

\begin{equation}T2_{\mu}\,^{\nu} = \tilde{R}_{\lambda \rho \mu \sigma}\tilde{R}^{\lambda \nu \rho \sigma}+\tilde{R}_{\lambda \rho \sigma \mu}\tilde{R}^{\lambda \sigma \rho \nu}-\frac{1}{2}\delta_{\mu}\,^{\nu}\tilde{R}_{\lambda \rho \tau \sigma}\tilde{R}^{\lambda \tau \rho \sigma}\,,\end{equation}

\begin{equation}T3_{\mu}\,^{\nu} = \tilde{R}_{\lambda \rho \mu \sigma}\tilde{R}^{\nu \sigma \lambda \rho}-\frac{1}{4}\delta_{\mu}\,^{\nu}\tilde{R}_{\lambda \rho \tau \sigma}\tilde{R}^{\tau \sigma \lambda \rho}\,,\end{equation}

\begin{equation}H1_{\mu}\,^{\nu} = \tilde{R}^{\nu}\,_{\lambda \mu \rho}\tilde{R}^{\lambda \rho}+\tilde{R}_{\lambda \mu}\tilde{R}^{\lambda \nu}-\frac{1}{2}\delta_{\mu}\,^{\nu}\tilde{R}_{\lambda \rho}\tilde{R}^{\lambda \rho}\,,\end{equation}

\begin{equation}H2_{\mu}\,^{\nu} = \tilde{R}^{\nu}\,_{\lambda \mu \rho}\tilde{R}^{\rho \lambda}+\tilde{R}_{\lambda \mu}\tilde{R}^{\nu \lambda}-\frac{1}{2}\delta_{\mu}\,^{\nu}\tilde{R}_{\lambda \rho}\tilde{R}^{\rho \lambda}\,,\end{equation}

\begin{equation}C1_{\mu}\,^{\lambda \nu} = \nabla_{\rho}\tilde{R}_{\mu}\,^{\lambda \rho \nu}+K^{\lambda}\,_{\sigma \rho}\tilde{R}_{\mu}\,^{\sigma \rho \nu}-K^{\sigma}\,_{\mu \rho}\tilde{R}_{\sigma}\,^{\lambda \rho \nu}\,,\end{equation}

\begin{equation}C2_{\mu}\,^{\lambda \nu} = \nabla_{\rho}\left(\tilde{R}_{\mu}\,^{\nu \lambda \rho}-\tilde{R}_{\mu}\,^{\rho \lambda \nu}\right)+K^{\lambda}\,_{\sigma \rho}\left(\tilde{R}_{\mu}\,^{\nu \sigma \rho}-\tilde{R}_{\mu}\,^{\rho \sigma \nu}\right)-K^{\sigma}\,_{\mu \rho}\left(\tilde{R}_{\sigma}\,^{\nu \lambda \rho}-\tilde{R}_{\sigma}\,^{\rho \lambda \nu}\right)\,,\end{equation}

\begin{equation}C3_{\mu}\,^{\lambda \nu} = \nabla_{\rho}\tilde{R}^{\rho \nu \lambda}\,_{\mu}+K^{\lambda}\,_{\sigma \rho}\tilde{R}^{\rho \nu \sigma}\,_{\mu}-K^{\sigma}\,_{\mu \rho}\tilde{R}^{\rho \nu \lambda}\,_{\sigma}\,,\end{equation}

\begin{equation}Y1_{\mu}\,^{\lambda \nu}=\delta_{\mu}\,^{\nu}\nabla_{\rho}\tilde{R}^{\lambda \rho}-\nabla_{\mu}\tilde{R}^{\lambda \nu}+\delta_{\mu}\,^{\nu}K^{\lambda}\,_{\sigma \rho}\tilde{R}^{\sigma \rho}+K^{\rho}\,_{\mu \rho}\tilde{R}^{\lambda \nu}-K^{\nu}\,_{\mu \rho}\tilde{R}^{\lambda \rho}-K^{\lambda}\,_{\rho \mu}\tilde{R}^{\rho \nu}\,,\end{equation}

\begin{equation}Y2_{\mu}\,^{\lambda \nu} = \delta_{\mu}\,^{\nu}\nabla_{\rho}\tilde{R}^{\rho \lambda}-\nabla_{\mu}\tilde{R}^{\nu \lambda}+\delta_{\mu}\,^{\nu}K^{\lambda}\,_{\sigma \rho}\tilde{R}^{\rho \sigma}+K^{\rho}\,_{\mu \rho}\tilde{R}^{\nu \lambda}-K^{\nu}\,_{\mu \rho}\tilde{R}^{\rho \lambda}-K^{\lambda}\,_{\rho \mu}\tilde{R}^{\nu \rho}\,.\end{equation}

It is worthwhile to stress that all these quantities have a tensor character induced by the nature of the curvature and the torsion tensors, so that the physics equations depending on them retain the same form independently of the choice of coordinates on the manifold, according to the principle of general covariance.

Then, the field equations are derived from the PG action by performing variations with respect to the gauge potentials:

\begin{equation}\delta S=\frac{1}{16 \pi}\int{\left(e_{a}\,^{\mu}X1_{\mu}\,^{\nu}\delta e^{a}\,_{\nu}+e_{a}\,^{\mu}e_{b \lambda}X2_{\mu}\,^{\lambda \nu}\delta \omega^{a b}\,_{\nu}\right)\sqrt{-g}\,d^4x}\,,\end{equation}
so that they constitute the following system of equations:

\begin{equation}X1_{\mu}\,^{\nu} = 0\,,\end{equation}

\begin{equation}\label{connectioneqq}X2_{[\mu \lambda]}\,^{\nu} = 0\,,\end{equation}
where:

\begin{equation}X1_{\mu}\,^{\nu} = -2G_{\mu}\,^{\nu}+4c_{1}T1_{\mu}\,^{\nu}+2c_{2}T2_{\mu}\,^{\nu}-2\left(2c_{1}+c_{2}\right)T3_{\mu}\,^{\nu}+2d_{1}\left(H1_{\mu}\,^{\nu}-H2_{\mu}\,^{\nu}\right)\,,\end{equation}

\begin{equation}\label{connectioneq}X2_{\mu}\,^{\lambda \nu} = 4c_{1}C1_{\mu}\,^{\lambda \nu}-2c_{2}C2_{\mu}\,^{\lambda \nu}+2\left(2c_{1}+c_{2}\right)C3_{\mu}\,^{\lambda \nu}-2d_{1}\left(Y1_{\mu}\,^{\lambda \nu}-Y2_{\mu}\,^{\lambda \nu}\right)\,.\end{equation}

On the other hand, the static spherically symmetric line element and the respective tetrad basis are chosen as:

\begin{equation}ds^2=\Psi_{1}(r)\,dt^2-\frac{dr^2}{\Psi_{2}(r)}-r^2\left(d\theta_{1}^{2}+\sin^{2}{\theta_{1}}d\theta_{2}^{2}\right)\,,\end{equation}

\begin{equation}
e^{\hat{t}}=\sqrt{\Psi_{1}(r)}\,dt\,,\;\;\;
e^{\hat{r}}=\frac{dr}{\sqrt{\Psi_{2}(r)}}\,,\;\;\;
e^{\hat{\theta_{1}}}=r\,d \theta_{1}\,,\;\;\;
e^{\hat{\theta_{2}}}=r\sin\theta_{1} \, d \theta_{2}\,;
\end{equation}
with $0 \leq \theta_{1} \leq \pi$ and $0 \leq \theta_{2} \leq 2\pi$.

In addition, torsion must satisfy the condition $\mathcal{L}_{\xi}T^{\lambda}\,_{\mu \nu}=0$ (i.e. the Lie derivative in the direction of the Killing vector $\xi$ on $T^{\lambda}\,_{\mu \nu}$ vanishes), in order to preserve the symmetry properties of the system. Then, the only non-vanishing components of $T^{\lambda}\,_{\mu \nu}$ are \cite{Rauch-Nieh,Sur-Bhatia}:

\begin{eqnarray}
T^{t}\,_{t r}&=&a(r) \;\;\, ,\nonumber\\
T^{r}\,_{t r}&=&b(r) \;\;\; ,\nonumber\\
T^{\theta_{k}}\,_{t \theta_{k}}&=&c(r) \;\;\; ,\nonumber\\
T^{\theta_{k}}\,_{r \theta_{k}}&=&g(r) \;\;\; ,\nonumber\\
T^{\theta_{k}}\,_{t \theta_{l}}&=&e^{a \theta_{k}}\,e^{b}\,_{\theta_{l}}\,\epsilon_{a b}\, d (r) \;\;\; , \nonumber\\
T^{\theta_{k}}\,_{r \theta_{l}}&=&e^{a \theta_{k}}\,e^{b}\,_{\theta_{l}}\,\epsilon_{a b}\, h (r) \;\;\; ,  \nonumber\\
T^{t}\,_{\theta_{k} \theta_{l}}&=&\epsilon_{k l} \, k (r)\,\sin\theta_{1} \;\;\; , \nonumber\\
T^{r}\,_{\theta_{k} \theta_{l}}&=&\epsilon_{k l} \, l (r)\,\sin\theta_{1} \;\;\; ;
\end{eqnarray}
where $a,b,c,d,g,h,k$ and $l$ are arbitrary functions depending only on r; $k,l=1,2$ with $k \neq l$ and $\epsilon_{a b}$ is the totally antisymmetric Levi-Civita symbol, given by:

\begin{equation}
\epsilon_{a b} = \left\{
\begin{array}{l}
+1\,, \;\;\;\; \text{for} \;\;\; a\,b=1\,2. \\
-1\,, \;\;\;\; \text{for} \;\;\; a\,b=2\,1. \\
\;\;\;0\,, \;\;\;\; \text{for all other combinations}.
\end{array}
\right.
\end{equation}

As can be seen, the SO(3)-symmetrical torsion exhibits eight degrees of freedom and it allows us to consider the most general expression for the torsion tensor. It means the possible existence of more complex solutions than the O(3)-symmetrical torsion case, where only four degrees of freedom survive.

Nevertheless, it is possible to impose an additional restriction involving these torsion components by taking the trace of Eq. (\ref{connectioneqq}) in the weak-field approximation:

\begin{eqnarray}
\left(4c_{1}+c_{2}+d_{1}\right)\nabla_{\rho}\tilde{R}^{[\lambda \rho]} &=& \; 2c_{1}K_{\mu \nu \rho}\left(\tilde{R}^{\nu \lambda \rho \mu}-\tilde{R}^{\rho \mu \nu \lambda}\right)+\frac{3}{2}c_{2}K_{\mu \nu \rho}\left(\tilde{R}^{\mu [\rho \nu \lambda]}+\tilde{R}^{\rho [\mu \lambda \nu]}\right)
\nonumber\\
&+&d_{1}\left(K_{\mu \rho}\,^{\lambda}\tilde{R}^{[\mu \rho]}+T^{\rho}\,_{\mu \rho}\tilde{R}^{[\lambda \mu]}\right)-\left(4c_{1}+c_{2}+d_{1}\right)K^{\lambda}\,_{\nu \rho}\tilde{R}^{[\nu \rho]} \, .
\end{eqnarray}

Then, by neglecting torsion terms of second order, only the first term of the equation contributes. The equations of motion for the torsion tensor in linear approximation read

\begin{equation}\nabla_{\mu}\nabla^{\mu}T^{\nu}\,_{\lambda \nu}+\nabla_{\mu}\nabla_{\nu}T^{\nu \mu}\,_{\lambda}-\nabla_{\mu}\nabla_{\lambda}T^{\nu \mu}\,_{\nu} = 0 \,,\end{equation}
for theories with $4c_{1}+c_{2}+d_{1} \neq 0$.

In terms of the torsion components, this constraint is equivalent to the relation:

\begin{equation}
\label{rel}
b(r)=rc\,'(r)+c(r)+\frac{p}{r}\,
\sqrt{\frac{\Psi_{1}(r)}{\Psi_{2}(r)}}\,,
\end{equation}
where $p$ is an integration constant.

In addition to a cosmological constant, we only focus on suitable solutions that may exist in presence of Coulomb electric and magnetic fields, as in the standard Einstein-Maxwell framework of GR, so the solutions are restricted to verify $\Psi_{1}(r) = \Psi_{2}(r)\equiv \Psi(r)$ in order to satisfy the Maxwell's equations in the RC manifold. These restrictions substantially simplify the problem. In any case, the field equations constitute a highly nonlinear system involving a large number of degrees of freedom and it forms an underdetermined system with different classes of solutions. We will require a final additional condition: suitable solutions must take an appropriate form referred to the rotated basis $\vartheta^{a}=\Lambda^{a}\,_{b}e^{b}$, given by the following vector fields:

\begin{eqnarray}
\vartheta^{\hat{t}}&=&\frac{1}{2}\left\{ \left[\Psi(r)+1\right]\,dt+\left[1-\frac{1}{\Psi(r)}\right]\,dr \right\} \;\; ;\nonumber\\
\vartheta^{\hat{r}}&=&\frac{1}{2}\left\{ \left[\Psi(r)-1\right]\,dt+\left[1+\frac{1}{\Psi(r)}\right]\,dr \right\} \;\; ;\nonumber\\
\vartheta^{\hat{\theta}_{1}}&=&r\,d \theta_{1} \;\; ;\nonumber\\%
\vartheta^{\hat{\theta_{2}}}&=&r\sin\theta_{1} \, d \theta_{2} \;\;.
\end{eqnarray}

This orthogonal coframe has already been used in previous literature to simplify the form of the Baekler solution, that belongs to a different class of PG models containing an O(3)-symmetrical torsion \cite{Baekler}. Especially, besides to its considerable simplification of the solution, it has the advantage of leading to a conformally flat Lorentz connection \cite{BaeklerHecht,GronHehl}. In our case, we expect that the rotated Lorentz connection defined on the RC manifold recovered its Minkowski values for the vanishing of the free parameters associated with the torsion tensor and then the remaining physical configuration reduced to GR. Note that, in order to reach this limit, it is not necessary that each component of torsion vanishes identically, but only the fulfillment of the first Bianchi identity of GR for the total curvature, as remarked in the previous section.

At the same time, any solution $F^{a}\,_{b c}$ referring to the mentioned orthogonal coframe can be written as follows:

\begin{eqnarray}
\mathcal{F}^{\hat{t}}\,_{\hat{t} \hat{r}} &=& \frac{1}{2}\left\{\left[1+\Psi(r)\right]a(r)+\left[1-\frac{1}{\Psi(r)}\right]b(r)\right\} \;\;\; ;\nonumber\\
\mathcal{F}^{\hat{r}}\,_{\hat{t} \hat{r}} &=& \frac{1}{2}\left\{\left[1+\frac{1}{\Psi(r)}\right]b(r)-\left[1-\Psi(r)\right]a(r)\right\} \;\;\; ;\nonumber\\
\mathcal{F}^{\hat{\theta_{1}}}\,_{\hat{t} \hat{\theta_{1}}} &=& \mathcal{F}^{\hat{\theta_{2}}}\,_{\hat{t} \hat{\theta_{2}}} = \frac{1}{2}\left\{\left[1+\frac{1}{\Psi(r)}\right]c(r)+\left[1-\Psi(r)\right]g(r)\right\} \;\;\; ;\nonumber\\
\mathcal{F}^{\hat{\theta_{1}}}\,_{\hat{r} \hat{\theta_{1}}} &=& \mathcal{F}^{\hat{\theta_{2}}}\,_{\hat{r} \hat{\theta_{2}}} = \frac{1}{2}\left\{\left[1+\Psi(r)\right]g(r)-\left[1-\frac{1}{\Psi(r)}\right]c(r)\right\} \;\;\; ;\nonumber\\
\mathcal{F}^{\hat{\theta_{2}}}\,_{\hat{t} \hat{\theta_{1}}} &=& - \mathcal{F}^{\hat{\theta_{1}}}\,_{\hat{t} \hat{\theta_{2}}} = \frac{1}{2}\left\{\left[1+\frac{1}{\Psi(r)}\right]d(r)+\left[1-\Psi(r)\right]h(r)\right\} \;\;\; ;\nonumber\\
\mathcal{F}^{\hat{\theta_{2}}}\,_{\hat{r} \hat{\theta_{1}}} &=& - \mathcal{F}^{\hat{\theta_{1}}}\,_{\hat{r} \hat{\theta_{2}}} = \frac{1}{2}\left\{\left[1+\Psi(r)\right]h(r)-\left[1-\frac{1}{\Psi(r)}\right]d(r)\right\} \;\;\; ;\nonumber\\
\mathcal{F}^{\hat{t}}\,_{\hat{\theta_{1}} \hat{\theta_{2}}} &=& \frac{1}{2r^2}\left\{\left[1+\Psi(r)\right]k(r)+\left[1-\frac{1}{\Psi(r)}\right]l(r)\right\} \;\;\; ;\nonumber\\
\mathcal{F}^{\hat{r}}\,_{\hat{\theta_{1}} \hat{\theta_{2}}} &=& \frac{1}{2r^2}\left\{\left[1+\frac{1}{\Psi(r)}\right]l(r)-\left[1-\Psi(r)\right]k(r)\right\} \;\;\; .
\end{eqnarray}

Therefore, in order to obtain a class of suitable non-singular solutions (excluding the point $r=0$), the components of the torsion tensor must satisfy the following relations:

\begin{equation}
\label{rel2}
b(r) = a(r)\,\Psi(r)\,,\;\;\;c(r) = - \, g(r)\,\Psi(r)\,,\;\;\;d(r) = - \, h(r)\,\Psi(r)\,,\;\;\;l(r) = k(r)\,\Psi(r)\,.
\end{equation}

We find out that these constraints also involve the vanishing of the three independent quadratic torsion invariants (i.e. $T_{\lambda \mu \nu}\,T^{\lambda \mu \nu}=T_{\lambda \mu \nu}\,T^{\mu \lambda \nu}=T^{\mu}\,_{\mu \lambda}\,T_{\nu}\,^{\nu \lambda}=0$).

\section{Solutions}

By taking into account all these remarks, the following SO(3)-symmetric vacuum solution can be easily found for $c_{1} = -\,d_{1}/4$ and $c_{2} = -\,d_{1}/2$:

\begin{eqnarray}
a(r)&=&\frac{\Psi'(r)}{2\Psi(r)}\,,\;\;
b(r)=\frac{\Psi'(r)}{2}\,,\;\;
c(r)=\frac{\Psi(r)}{2r}\,,\;\;
g(r)=-\,\frac{1}{2r}\,,\;\;
d(r)=\frac{\kappa}{r}\,,\;\;
h(r)=-\,\frac{\kappa}{r\Psi(r)}\,,\;\;
k(r)=l(r)=0\;;
\end{eqnarray}
with

\begin{eqnarray}
\Psi(r)&=&1-\frac{2m}{r}+\frac{d_{1}\kappa^{2}}{r^2} \;\;.
\end{eqnarray}

Hence, the relation (\ref{rel}) is completely fulfilled and the constant $p$ vanishes.

This solution describes a Reissner-Nordstr\"{o}m type geometry, supported only by the metric and torsion fields rather than an electric or magnetic source. The new contribution is proportional to the square of the new parameter $\kappa$.
Indeed, this parameter determines the intensity of the strength tensor corresponding to the torsion:

\begin{eqnarray}
F^{a b}\,_{c d} &=& \left(
\begin{array}{cccccc}
0 & 0 & 0 & 0 & 0 & 0 \\
0 & 0 & -\,\kappa/2r^2 & 0 & -\,\kappa/2r^2 & 0 \\
0 & \kappa/2r^2 & 0 & 0 & 0 & -\,\kappa/2r^2 \\
-\,\kappa/r^2 & 0 & 0 & -\,1/r^2 & 0 & 0 \\
0 & \kappa/2r^2 & 0 & 0 & 0 & -\,\kappa/2r^2 \\
0 & 0 & \kappa/2r^2 & 0 & \kappa/2r^2 & 0
\end{array} \right)\,,
\end{eqnarray}
where the six rows and columns of the matrix are labeled the components in the order (01, 02, 03, 23, 31, 12).

The values above for the Lagrangian coefficients and their respective signs define the strength and properties of the torsion field in the PG framework. In existing literature, particular results containing a certain set of viable coefficient combinations for the purely massless PG theory have been developed under the linear field approximation requiring the absence of both ghost and tachyon modes \cite{Kuhfuss} or only the ghost-free condition \cite{Sezgin,Fukuma}. Nevertheless, it has also been shown that the Hamiltonian constraint formalism differs from these results where the highly nonlinear effects of the PG theory are included \cite{Yo}. Furthermore, some other authors have pointed out several mistakes and incompleteness in various of the mentioned analyses, reaching important contradictions with the commented conclusions \cite{Blagojevic,Battiti}. In this sense, the stability of these models is still an open issue.

On the other hand, by following our constraints (\ref{rel}) and (\ref{rel2}), we note that any other combination for the constant parameters of Eq. \eqref{actioneq} involves a vacuum configuration described strictly by the Schwarzschild metric. Hence, in the present case, there is a unique combination that allows a vacuum configuration different from the Schwarzschild geometry. It is the Reissner-Nordstr\"{o}m solution above.
Moreover, by solving the field equations it is possible to demonstrate this statement even for the case $\Psi_1(r)\neq\Psi_2(r)$.
It is also shown that the torsion decreases at infinity and the metric is asymptotically flat. So the corresponding Newtonian limit is satisfied by the solution as demanded by different approaches \cite{Chen}.

It is also straightforward to notice that the condition $\tilde{\nabla}_{[\mu}T^{\lambda}\,_{\nu \rho]}+T^{\sigma}\,_{[\mu \nu}\,T^{\lambda}\,_{\rho] \sigma}=0$ is fulfilled for this solution when $\kappa=0$. In such a case, although the rest of the non-vanishing components of the torsion tensor still remain, the Action \eqref{actioneqq} is equivalent to the Einstein-Hilbert one and the GR approach is totally recovered. These non-vanishing components yield an inert RC spin connection and curvature, which emerge to the physical structure only when the parameter $\kappa$ switches on and the torsion becomes dynamical. This fact contrasts with the alternative ways of recovering the regular gravity action given by the rest of the PG models present in previous literature, such as the mentioned Baekler solution where this limit is carried out in the framework of teleparallelism \cite{Hehl:1978yt}. Teleparallel Gravity is the gauge theory for the translation group based on the curvature-free Weitzenb\"{o}ck connection and it is constructed in such a form that provides an equivalent description of gravity to GR, but in terms of torsion so that there exist conceptual differences between them (see \cite{Cai} for a recent overview).

Additionally, the expression for the Lorentz connection referred to $\vartheta^{a}$ exhibits a similar property to its counterpart of the Baekler solution. It takes Minkowski values within the RC manifold for $\kappa=0$ and it does not depend on any other magnitude in such a case:

\begin{equation}
\hat{A}=
-\,\frac{\kappa}{r}\,J_{\hat{\theta}\hat{\phi}}\,dt
+\frac{\kappa}{r\Psi(r)}\,J_{\hat{\theta}\hat{\phi}}\,dr
+\frac{1}{2}\left(J_{\hat{r} \hat{\theta}}-J_{\hat{t} \hat{\theta}}\right)\,d\theta +\sin \theta \left[\frac{1}{2}\left(J_{\hat{r} \hat{\phi}}-J_{\hat{t} \hat{\phi}}\right)+\cot \theta J_{\hat{\theta} \hat{\phi}} \right]\,d\phi\,.
\end{equation}

This solution can be trivially generalized to include the existence of a non-vanishing cosmological constant $\Lambda$ and Coulomb electromagnetic fields produced by both electric and magnetic charges $q_{e}$ and $q_{m}$, respectively. For this purpose, it is assumed that photons are decoupled from torsion as it is dictated by the minimum coupling principle. Then, it is easy to extend the solution by modifying the metric function $\Psi(r)$ by the following expression:

\begin{equation}\Psi(r)=1-\frac{2m}{r}+\frac{d_{1}\kappa^{2}+q_{e}^{2}+q_{m}^{2}}{r^2}+\frac{\Lambda}{3}r^2\,.\end{equation}

As can be seen, the term derived by the dynamical torsion has the same structure than the terms provided by the electric and magnetic monopole charges and it is possible to collect these three contributions along with the cosmological constant onto a common space-time. Therefore, these factors involve geometrical effects on the PG field strength tensors, even though the electromagnetic field is not coupled directely to the torsion field. Switching off the parameter $\kappa$, the solution reduces to the Reissner-Nordstr\"{o}m-de Sitter solution of ordinary GR as expected.
Thereby, this solution shows similarities between the torsion and the electromagnetic fields, even though they are independent quantities.

It is worthwhile to stress the further relation between this type of geometry and other well known post-Riemannian approaches, such as the metric-affine gauge (MAG) theory of gravity, where the RC space-time and the PG group are both replaced by a general affinely connected metric manifold with non-metricity condition (i.e. $\tilde{\nabla}_{\lambda}\,g_{\mu\nu} \neq 0$) and its associated affine gauge group \cite{HHehl,Blag-Hehl}. Indeed, analogous results were found out in terms of the dilation and the shear charges associated with the non-metricity tensor, which can involve a vacuum Reissner-Nordstr\"{o}m configuration in this context \cite{Tresguerres,Hehl-Mac}. Nevertheless, the so called gravito-electric and gravito-magnetic terms present in all these solutions fall completely on the non-metricity field, so that when the latter vanishes those terms disappear from the metric tensor, even in presence of a non-vanishing torsion component. This result differs from our PG solution since the Reissner-Nordstr\"{o}m structure provided by the torsion field can even exist when the connection is metric-compatible and the non-metricity tensor vanishes. This fact together with the mentioned achievements of the MAG point out a richer structure of spherical and static solutions in gravitational theories characterized by a general affine connection.

\section{Equations of motion}

As any test particle or physical field uncoupled to torsion cannot experiment deviations from their geodesic trajectories, the respective equations of motion within the RC space-time connected to our PG model must distinguish between both classes of spinless and spinning matter.
For this purpose, it is critical to deal with the principal conservation law of the total energy-momentum tensor $\theta^{\mu \nu}$ derived by the invariance of Action (\ref{actioneqq}):

\begin{equation}
\nabla_{\nu}\theta^{\mu \nu} + K_{\lambda \rho}\,^{\mu}\theta^{\rho \lambda} + \tilde{R}_{\lambda \rho \sigma}\,^{\mu}\,S^{\lambda \rho \sigma}=0\,,
\end{equation}
where $S^{\lambda \rho \sigma}$ is the spin density tensor.

An analysis for the achievement of this result based on our particular PG model is shown in the Appendix A. The mentioned conservation law allows to obtain the equations of motion for a test particle in such a RC space-time by integrating the expression above over a three dimensional space-like section of the world tube involving the particle and employing the semiclassical approximation \cite{Papapetrou,Hehll}:

\begin{equation}
\label{Ineq}
\int{\partial_{\nu}\left(\sqrt{-g}\,\theta^{\mu \nu}\right)d^{3}x'}+\int{\Gamma^{\mu}\,_{\lambda \rho}\theta^{\lambda \rho}\sqrt{-g}\,d^{3}x'}+\int{K_{\lambda \rho}\,^{\mu}\theta^{\rho \lambda}\sqrt{-g}\,d^{3}x'}+\int{\tilde{R}_{\lambda \rho \sigma}\,^{\mu}\,S^{\lambda \rho \sigma}\sqrt{-g}\,d^{3}x'}=0,
\end{equation}
with

\begin{equation}
\int{\partial_{\nu}
\left(\sqrt{-g}\,\theta^{\mu \nu}\right)
d^{3}x'}
=
\frac{d}{dt}
\int{\theta^{\mu t}\sqrt{-g}\,d^{3}x'}\,,
\end{equation}
due to the Gauss theorem and by neglecting surface terms. As Eq. \eqref{Ineq} must be
fulfilled for any integration volume, it is equivalent to the differential
equation of motion:

\begin{equation}
\frac{dp^{\mu}}{ds}+\Gamma^{\mu}\,_{\lambda \rho}\,p^{\lambda}u^{\rho}+K_{\lambda \rho}\,^{\mu}p^{\rho}u^{\lambda}+\tilde{R}_{\lambda \rho \sigma}\,^{\mu}S^{\lambda \rho}u^{\sigma}=0\,,
\end{equation}
where we have used the following definitions

\begin{equation}
\theta^{\lambda \rho}=\frac{dt}{ds}\int{p^{\lambda}u^{\rho}\sqrt{-g}\,d^{3}x'}\,,
\end{equation}
and

\begin{equation}
S^{\lambda \rho \sigma}=\frac{dt}{ds}\int{S^{\lambda \rho}u^{\sigma}\sqrt{-g}\,d^{3}x'}\,.
\end{equation}

Here, $s$ is the proper time along the particle world line, $p^{\mu}$ the four-momentum of the particle and $u^{\mu}$ its four-velocity.
Therefore, the presence of a dynamical torsion in the space-time and the interaction between the curvature and the spin of matter originate in general, a generalized Lorentz force acting on this type of matter. Thus, this force potentially yields deviations from the geodesic trajectories. Of course, this generally non-geodesic motion turns out to be another essential difference with gravitational theories endowed with vanishing torsion, such as ordinary GR. Nevertheless, for spinless matter with $S^{\lambda \rho}=0$ and $p^{\lambda} \propto u^{\lambda}$, the equations of motion reduce to the same geodesic equations of GR.

This fundamental difference might be used in order to prove experimentally the possible existence of a non-vanishing dynamical torsion in the space-time. Nevertheless, it is expected to yield too tiny effects to be measured, as occurs with the rest of the well known PG models. Additionally, torsion is induced on the vierbein field by the field equations and thereby it can also operate on the geodesic motion of ordinary matter via the Levi-Civita connection. In particular, for a standard Reissner-Nordstr\"{o}m geometry, the respective point charges have well known consequences on the geodesic paths of test particles and light rays \cite{Hackmann}.

Presumably, the effects of this type of geometry are also very small at astrophysics or cosmological scales, because of the vanishing of the spin density tensor in the most macroscopical bodies. However, this situation may differ around extreme gravitational systems as neutron stars or black holes with intense magnetic fields and sufficiently oriented elementary spins. In such a case, it is expected that the RC space-time described by the PG theory modulates these events.

Further analyses can be performed by comparing the gravitational interaction of the spin and the orbital angular momentum of a rotating rigid test body \cite{Stoeger,Yasskin}. In this sense, it is especially interesting their natural extension towards the MAG theory when the motion of a rotating and deformable test body is considered \cite{Puetzfeld}. All these achievements allow to systematically study the behaviour of gravitating matter with microstructure and to establish additional differences between a large extreme gravitational systems, such as the one present in our PG model and the one previously mentioned supported by MAG.

\section{Conclusions}

In the present work, we have investigated the PG theory with massless torsion based on a gravitational model directly connected to GR when the dynamical role of torsion is frozen via the first Bianchi identity. In the general case, this approach allows the torsion tensor to constitute a dynamical degree of freedom. We have shown that the vacuum structure of the theory may differ from the Einstein's theory and, specifically, distinct classes of solutions can exist besides the Schwarzschild solution given by the Birkhoof's theorem within the standard framework of GR. Hence, in order to improve the understanding of such a theory of gravity, the search and analysis of exact solutions are fundamental.

The large degree of symmetry assumed and the requirement of the existence of a suitable electromagnetic-like vacuum structure analogous to the Einstein-Maxwell framework together with the use of a convenient rotated basis allow to reduce notably the difficulty of the highly nonlinear nature present in the theory. Under these requirements, we have obtained a new static and spherically symmetric vacuum solution. This solution provides a Reissner-Nordstr\"{o}m type geometry with a SO(3)-symmetrical torsion depending on a parameter $\kappa$ and it has been deduced without the use of the double duality ansatz for the RC curvature, often employed in previous literature in order to restrict the PG field equations into a very highly simplified system \cite{Mielke}. Its existence shows the dynamical character of the torsion field, which can even be induced on the metric tensor via the field equations generating a distinct class of solutions, beyond the Schwarzschild scheme and the Birkhoff's theorem of GR.

The corresponding generalized Reissner-Nordstr\"{o}m-de Sitter configuration is also obtained when external electromagnetic fields and a non-vanishing cosmological constant are included, by analogy with the standard case. In this scheme, the torsion field contribution is perfectly distinguishable from the rest of physical degrees of freedom and the solution reduces to the standard case when its dynamical role is switched off. Therefore, the solution presents similarities between the torsion and the electromagnetic fields. It is expected that these similarities still remain in more general systems, such as axisymmetric space-times.

The foundations presented in this article have also been employed in previous works for the analysis and the achievement of exact solutions in extended models of gravity, such as the well known Einstein-Yang-Mills theory. The results obtained in this work show the flexibility and usefulness of the method described in \cite{CembGig,CembGigg}. Furthermore, the recurrence of the fundamental schemes derived by our analyses in the extensive MAG framework is also remarked. It shows deeper relations between the solutions and the vacuum structure provided by these approaches, which improve their physical understanding and applicability. Specifically, the role of the non-metricity present in MAG has been typically categorized into earlier eppochs of the universe, whereas the one of the torsion field is expected to represent a larger number of physical scenarios, even in our current universe, such as extreme gravitational systems described by neutron stars or black holes with intense spin densities.

Finally, the equations of motion for a general test particle are derived and the differences with the geodesic trajectories of GR are stressed. These differences are also very important to understand the physical properties and further implications of our solution. Their theoretical consequences or observational effects in astrophysics and cosmology will be studied in future work.

\section{Appendix A}

The conservation law for the total energy-momentum tensor associated with our model can be obtained directly from the PG Lagrangian:

\begin{equation}
S=\frac{1}{16 \pi}
\int{d^4x\,\sqrt{-g}
\left\{
\mathcal{L}_{m}-{R}+\frac{d_{1}}{4}
\left[
2\tilde{R}_{\lambda \rho \mu \nu}\tilde{R}^{\mu \nu \lambda \rho}-\tilde{R}_{\lambda \rho \mu \nu}\tilde{R}^{\lambda \rho \mu \nu}-2\tilde{R}_{\lambda \rho \mu \nu}\tilde{R}^{\lambda \mu \rho \nu}+4\tilde{R}_{\mu \nu}
\left(
\tilde{R}^{\mu \nu}-\tilde{R}^{\nu \mu}
\right)
\right]
\right\}
}\,.
\end{equation}

We can obtain this result by the computation of the torsion-free divergence acting on the vierbein equation:

\begin{eqnarray}
\nabla_{\nu}X1_{\mu}\,^{\nu} &=& \; d_{1}\left[\tilde{R}_{\lambda \rho \mu \sigma}\left(\nabla_{\nu}\tilde{R}^{\lambda \sigma \rho \nu}-\nabla_{\nu}\tilde{R}^{\lambda \rho \nu \sigma}-\nabla_{\nu}\tilde{R}^{\lambda \nu \rho \sigma}+2\nabla_{\nu}\tilde{R}^{\nu \sigma \lambda \rho}\right)+2\left(\tilde{R}^{\lambda \nu}-\tilde{R}^{\nu \lambda}\right)\nabla_{\nu}\tilde{R}_{\lambda \mu}\right.  \nonumber\\
&+& \frac{1}{2}\tilde{R}_{\lambda \rho \omega \sigma}\left(\nabla_{\mu}\tilde{R}^{\lambda \rho \omega \sigma}+2\nabla_{\mu}\tilde{R}^{\lambda \omega \rho \sigma}-2\nabla_{\mu}\tilde{R}^{\omega \sigma \lambda \rho}\right)-2\left(\tilde{R}^{\lambda \rho}-\tilde{R}^{\rho \lambda}\right)\nabla_{\mu}\tilde{R}_{\lambda \rho}+2\left(\tilde{R}^{\lambda \rho}-\tilde{R}^{\rho \lambda}\right)\nabla_{\nu}\tilde{R}^{\nu}\,_{\lambda \mu \rho}  \nonumber\\
&+& \left. 2\tilde{R}^{\nu}\,_{\lambda \mu \rho}\nabla_{\nu}\left(\tilde{R}^{\lambda \rho}-\tilde{R}^{\rho \lambda}\right)+2\tilde{R}_{\lambda \mu}\nabla_{\nu}\left(\tilde{R}^{\lambda \nu}-\tilde{R}^{\nu \lambda}\right)+\nabla_{\nu}\tilde{R}_{\lambda \rho \mu \sigma}\left(\tilde{R}^{\lambda \sigma \rho \nu}-\tilde{R}^{\lambda \rho \nu \sigma}-\tilde{R}^{\lambda \nu \rho \sigma}+2\tilde{R}^{\nu \sigma \lambda \rho}\right)\right]\,.\nonumber\\
\end{eqnarray}

The information of the additional field equation $X2_{[\mu \lambda]}\,^{\nu}= - \, 16\pi S_{\lambda \mu}\,^{\nu}$, can be introduced in the equation above  with the result:

\begin{eqnarray}
\nabla_{\nu}X1_{\mu}\,^{\nu} &=& \; 16\pi \tilde{R}_{\lambda \rho \sigma \mu}\,S^{\lambda \rho \sigma}+ d_{1}
\left\{
\tilde{R}^{\lambda}\,_{\rho \mu \sigma}
\left[
K^{\rho}\,_{\omega \nu}\left(\tilde{R}_{\lambda}\,^{\omega \nu \sigma}+2\tilde{R}^{\nu \sigma \omega}\,_{\lambda}+\tilde{R}_{\lambda}\,^{\nu \omega \sigma}-\tilde{R}_{\lambda}\,^{\sigma \omega \nu}\right)\right.\right.  \nonumber\\
&-& K^{\omega}\,_{\lambda \nu}\left(\tilde{R}_{\omega}\,^{\rho \nu \sigma}+2\tilde{R}^{\nu \sigma \rho}\,_{\omega}+\tilde{R}_{\omega}\,^{\nu \rho \sigma}-\tilde{R}_{\omega}\,^{\sigma \rho \nu}\right) +2\delta^{\sigma}_{\lambda}\nabla_{\nu}\left(\tilde{R}^{\rho \nu}-\tilde{R}^{\nu \rho}\right)-2\nabla_{\lambda}\left(\tilde{R}^{\rho \sigma}-\tilde{R}^{\sigma \rho}\right)  \nonumber\\
&+& \left.  2\delta^{\sigma}_{\lambda}K^{\rho}\,_{\omega \nu}\left(\tilde{R}^{\omega \nu}-\tilde{R}^{\nu \omega}\right)+2K^{\nu}\,_{\lambda \nu}\left(\tilde{R}^{\rho \sigma}-\tilde{R}^{\sigma \rho}\right)-2K^{\sigma}\,_{\lambda \nu}\left(\tilde{R}^{\rho \nu}-\tilde{R}^{\nu \rho}\right)-2K^{\rho}\,_{\nu \lambda}\left(\tilde{R}^{\nu \sigma}-\tilde{R}^{\sigma \nu}\right)
\right]
\nonumber\\
&+& \nabla_{\nu}\tilde{R}_{\lambda \rho \mu \sigma}\left(\tilde{R}^{\lambda \sigma \rho \nu}-\tilde{R}^{\lambda \rho \nu \sigma}-\tilde{R}^{\lambda \nu \rho \sigma}+2\tilde{R}^{\nu \sigma \lambda \rho}\right)+\frac{1}{2}\nabla_{\mu}\tilde{R}_{\lambda \rho \nu \sigma}\left(\tilde{R}^{\lambda \rho \nu \sigma}+2\tilde{R}^{\lambda \nu \rho \sigma}-2\tilde{R}^{\nu \sigma \lambda \rho}\right)  \nonumber\\
&-& 2\left(\tilde{R}^{\lambda \rho}-\tilde{R}^{\rho \lambda}\right)\nabla_{\mu}\tilde{R}_{\lambda \rho}+2\left(\tilde{R}^{\lambda \rho}-\tilde{R}^{\rho \lambda}\right)\nabla_{\nu}\tilde{R}^{\nu}\,_{\lambda \mu \rho}+2\tilde{R}^{\nu}\,_{\lambda \mu \rho}\nabla_{\nu}\left(\tilde{R}^{\lambda \rho}-\tilde{R}^{\rho \lambda}\right)
\nonumber\\
&+& \left.  2\tilde{R}_{\lambda \mu}\nabla_{\nu}\left(\tilde{R}^{\lambda \nu}-\tilde{R}^{\nu \lambda}\right)+2\left(\tilde{R}^{\lambda \nu}-\tilde{R}^{\nu \lambda}\right)\nabla_{\nu}\tilde{R}_{\lambda \mu}
\right\}\,,
\end{eqnarray}
where

\begin{equation}
\tilde{R}^{\lambda}\,_{\rho \mu \sigma}\delta^{\sigma}_{\lambda}\nabla_{\nu}\left(\tilde{R}^{\rho \nu}-\tilde{R}^{\nu \rho}\right)+\tilde{R}_{\lambda \mu}\nabla_{\nu}\left(\tilde{R}^{\lambda \nu}-\tilde{R}^{\nu \lambda}\right)=\tilde{R}^{\nu}\,_{\lambda \mu \rho}\nabla_{\nu}\left(\tilde{R}^{\lambda \rho}-\tilde{R}^{\rho \lambda}\right)-\tilde{R}^{\lambda}\,_{\rho \mu \sigma}\nabla_{\lambda}\left(\tilde{R}^{\rho \sigma}-\tilde{R}^{\sigma \rho}\right)=0\,.
\end{equation}

First, we focus on the differential form of Riemann tensors and express the torsion-free operator $\nabla$ in terms of $\tilde{\nabla}$ and the contortion tensor:

\begin{equation}
\nabla_{\sigma}\tilde{R}_{\lambda \rho \mu \nu}=\tilde{\nabla}_{\sigma}\tilde{R}_{\lambda \rho \mu \nu}+K^{\omega}\,_{\lambda \sigma}\tilde{R}_{\omega \rho \mu \nu}+K^{\omega}\,_{\rho \sigma}\tilde{R}_{\lambda \omega \mu \nu}+K^{\omega}\,_{\mu \sigma}\tilde{R}_{\lambda \rho \omega \nu}+K^{\omega}\,_{\nu \sigma}\tilde{R}_{\lambda \rho \mu \omega}\,.
\end{equation}

Thus, by simplifying the resulting expression and rearranging terms, we obtain the following equation:

\begin{eqnarray}
\nabla_{\nu}X1_{\mu}\,^{\nu} &=& d_{1}
\left\{
\tilde{\nabla}_{\nu}\tilde{R}_{\lambda \rho \mu \sigma}\left(\tilde{R}^{\lambda \sigma \rho \nu}-\tilde{R}^{\lambda \rho \nu \sigma}-\tilde{R}^{\lambda \nu \rho \sigma}+2\tilde{R}^{\nu \sigma \lambda \rho}\right)+\frac{1}{2}\tilde{\nabla}_{\mu}\tilde{R}_{\lambda \rho \nu \sigma}\left(\tilde{R}^{\lambda \rho \nu \sigma}+2\tilde{R}^{\lambda \nu \rho \sigma}-2\tilde{R}^{\nu \sigma \lambda \rho}\right)\right.  \nonumber\\
&+& 2\tilde{R}^{\lambda}\,_{\rho \mu \sigma}
\left[
\delta^{\sigma}_{\lambda}K^{\rho}\,_{\omega \nu}\left(\tilde{R}^{\omega \nu}-\tilde{R}^{\nu \omega}\right)+K^{\nu}\,_{\lambda \nu}\left(\tilde{R}^{\rho \sigma}-\tilde{R}^{\sigma \rho}\right)-K^{\sigma}\,_{\lambda \nu}\left(\tilde{R}^{\rho \nu}-\tilde{R}^{\nu \rho}\right)-K^{\rho}\,_{\nu \lambda}\left(\tilde{R}^{\nu \sigma}-\tilde{R}^{\sigma \nu}\right)
\right]
\nonumber\\
&+& \frac{1}{2}\left(K^{\omega}\,_{\lambda \mu}\tilde{R}_{\omega \rho \nu \sigma}+K^{\omega}\,_{\rho \mu}\tilde{R}_{\lambda \omega \nu \sigma}+K^{\omega}\,_{\nu \mu}\tilde{R}_{\lambda \rho \omega \sigma}+K^{\omega}\,_{\sigma \mu}\tilde{R}_{\lambda \rho \nu \omega}\right)\left(\tilde{R}^{\lambda \rho \nu \sigma}+2\tilde{R}^{\lambda \nu \rho \sigma}-2\tilde{R}^{\nu \sigma \lambda \rho}\right)  \nonumber\\
&+& \left(K^{\omega}\,_{\mu \nu}\tilde{R}_{\lambda \rho \omega \sigma}+K^{\omega}\,_{\sigma \nu}\tilde{R}_{\lambda \rho \mu \omega}\right)\left(\tilde{R}^{\lambda \sigma \rho \nu}-\tilde{R}^{\lambda \rho \nu \sigma}-\tilde{R}^{\lambda \nu \rho \sigma}+2\tilde{R}^{\nu \sigma \lambda \rho}\right) + 2\left(\tilde{R}^{\lambda \rho}-\tilde{R}^{\rho \lambda}\right)\nabla_{\nu}\tilde{R}^{\nu}\,_{\lambda \mu \rho} \nonumber\\
&-&\left.  2\left(\tilde{R}^{\lambda \rho}-\tilde{R}^{\rho \lambda}\right)\nabla_{\mu}\tilde{R}_{\lambda \rho}+2\left(\tilde{R}^{\lambda \rho}-\tilde{R}^{\rho \lambda}\right)\nabla_{\rho}\tilde{R}_{\lambda \mu}
\right\}
+ 16\pi \tilde{R}_{\lambda \rho \sigma \mu}\,S^{\lambda \rho \sigma}\,.
\end{eqnarray}

According to the second Bianchi identity for a RC manifold, the components of the Riemann tensor satisfy \cite{Ortin}:

\begin{equation}
\tilde{\nabla}_{[\lambda |}\tilde{R}^{\sigma}\,_{\rho | \mu \nu]}-T^{\omega}\,_{[\lambda \mu |}\tilde{R}^{\sigma}\,_{\rho \omega | \nu]}=0\,,
\end{equation}
so that we can simplify even more terms and obtain the following expression:

\begin{eqnarray}
\nabla_{\nu}X1_{\mu}\,^{\nu} &=& d_{1}
\Bigl\{
\frac{1}{2}\left(K^{\omega}\,_{\rho \mu}\tilde{R}_{\lambda \omega \nu \sigma}-K^{\omega}\,_{\lambda \mu}\tilde{R}_{\rho \omega \nu \sigma}\right)\left(\tilde{R}^{\lambda \rho \nu \sigma}+2\tilde{R}^{\lambda \nu \rho \sigma}-2\tilde{R}^{\nu \sigma \lambda \rho}\right)-2\left(\tilde{R}^{\lambda \rho}-\tilde{R}^{\rho \lambda}\right)\nabla_{\mu}\tilde{R}_{\lambda \rho}\Bigr.  \nonumber\\
&+& 2\left(\tilde{R}^{\lambda \rho}-\tilde{R}^{\rho \lambda}\right)\left(\nabla_{\nu}\tilde{R}^{\nu}\,_{\lambda \mu \rho}-K^{\nu}\,_{\lambda \sigma}\tilde{R}^{\sigma}\,_{\nu \mu \rho}+K^{\omega}\,_{\nu \lambda}\tilde{R}^{\nu}\,_{\rho \mu \omega}-K^{\nu}\,_{\sigma \nu}\tilde{R}^{\sigma}\,_{\rho \mu \lambda}+\nabla_{\rho}\tilde{R}_{\lambda \mu}-\tilde{R}_{\nu \mu}K^{\nu}\,_{\lambda \rho}\right)
\nonumber\\
&+& \frac{1}{2}\left(\tilde{R}_{\lambda \rho \omega (\nu |}T_{\mu}\,^{\omega}\,_{| \sigma)}+\tilde{R}_{\lambda \rho \omega (\nu}T^{\omega}\,_{\sigma ) \mu}+\tilde{R}_{\lambda \rho \omega (\nu}T_{\sigma )}\,^{\omega}\,_{\mu}-\tilde{\nabla}_{(\nu |}\tilde{R}_{\lambda \rho \mu | \sigma )}\right)\left(\tilde{R}^{\lambda \rho \nu \sigma}-2\tilde{R}^{\nu \sigma \lambda \rho}\right)
\nonumber\\
&+&\Bigl. \tilde{R}_{\lambda \rho \mu \omega}
\left[
T^{\omega}\,_{\nu \sigma}\tilde{R}^{\lambda (\nu \sigma ) \rho}+2T_{(\nu \sigma)}\,^{\omega}\tilde{R}^{\lambda [\nu \sigma ] \rho}+T_{(\nu \sigma)}\,^{\omega}\left(2\tilde{R}^{\nu \sigma \lambda \rho}-\tilde{R}^{\lambda \rho \nu \sigma}\right)
\right]
\Bigr\}
+ 16\pi \tilde{R}_{\lambda \rho \sigma \mu}\,S^{\lambda \rho \sigma}\,.
\end{eqnarray}

The last factors vanish because of the contraction between the symmetric and antisymmetric parts of the tensors above. Then, by repeating the same procedure on the Ricci tensors:

\begin{eqnarray}
\nabla_{\nu}X1_{\mu}\,^{\nu} &=& d_{1}
\Bigl[
\frac{1}{2}\left(K^{\omega}\,_{\rho \mu}\tilde{R}_{\lambda \omega \nu \sigma}-K^{\omega}\,_{\lambda \mu}\tilde{R}_{\rho \omega \nu \sigma}\right)\left(\tilde{R}^{\lambda \rho \nu \sigma}+2\tilde{R}^{\lambda \nu \rho \sigma}-2\tilde{R}^{\nu \sigma \lambda \rho}\right)+2K^{\omega}\,_{\lambda \mu}\tilde{R}^{\rho \lambda}\tilde{R}_{\omega \rho}
\Bigr.
\nonumber\\
&+&
\Bigl.
2\left(\tilde{R}^{\lambda \rho}-\tilde{R}^{\rho \lambda}\right)K^{\omega}\,_{\nu \mu}\tilde{R}^{\nu}\,_{\lambda \omega \rho}+8\tilde{R}^{[\lambda \rho ]}K^{\omega}\,_{\nu (\lambda}\tilde{R}^{\nu}\,_{\rho ) \mu \omega}
\Bigr]
+ 16\pi \tilde{R}_{\lambda \rho \sigma \mu}\,S^{\lambda \rho \sigma}\,,
\end{eqnarray}
where, once again, the contraction $\tilde{R}^{[\lambda \rho ]}K^{\omega}\,_{\nu (\lambda}\tilde{R}^{\nu}\,_{\rho ) \mu \omega}=0$. On the other hand, the antisymmetric part of the energy-momentum tensor is related via the vierbein equation to the following quantity:
\begin{eqnarray}
X1 ^{[ \mu \nu ]} &=& \; \frac{d_{1}}{2}
\Bigl[
\tilde{R}_{\lambda \rho}\,^{\nu}\,_{\sigma}\left(\tilde{R}^{\lambda \mu \rho \sigma}-2\tilde{R}^{\mu \sigma \lambda \rho}\right)-\tilde{R}^{\nu}\,_{\sigma \lambda \rho}\left(\tilde{R}^{\rho \sigma \lambda \mu}-2\tilde{R}^{\lambda \rho \mu \sigma}\right)+2\left(\tilde{R}^{\mu \lambda}\tilde{R}_{\lambda}\,^{\nu}-\tilde{R}^{\lambda \mu}\tilde{R}^{\nu}\,_{\lambda}\right)
\Bigr.
\nonumber\\
&+&
\Bigl.
2\left(\tilde{R}_{\lambda \rho}-\tilde{R}_{\rho \lambda}\right)\left(\tilde{R}^{\nu \lambda \mu \rho}-\tilde{R}^{\mu \lambda \nu \rho}\right)
\Bigr]\,.
\end{eqnarray}

Therefore, it is straightforward to express this torsion-free divergence into a very concise form:

\begin{equation}
\nabla_{\nu}X1_{\mu}\,^{\nu} = K_{\lambda \rho \mu}X1^{\lambda \rho} + 16\pi \tilde{R}_{\lambda \rho \sigma \mu}\,S^{\lambda \rho \sigma}\,,
\end{equation}
and the general conservation law of the total energy-momentum tensor states from the equation $X1^{\mu \nu} = - \, 16 \pi \theta^{\mu \nu}$ in the following way:

\begin{equation}
\nabla_{\nu}\theta_{\mu}\,^{\nu} + K_{\lambda \rho \mu}\theta^{\rho \lambda} + \tilde{R}_{\lambda \rho \sigma \mu}\,S^{\lambda \rho \sigma}=0\,.
\end{equation}

\bigskip
\bigskip
\noindent
{\bf ACKNOWLEDGMENTS}

\bigskip
We would like to thank Prado Martin Moruno and Teodor Borislav Vasilev for helpful discussions.
This work has been supported in part by the MINECO (Spain) projects FIS2014-52837-P, FPA2014-53375-C2-1-P, and Consolider-Ingenio MULTIDARK CSD2009-00064.
J.A.R.C. acknowledges financial support from the {\it Jose Castillejo award (2015)}.

\end{document}